\documentclass[11pt]{iopart}

\usepackage{amsopn}
\usepackage{iopams}
\usepackage{graphicx}
\usepackage[english]{babel}
\usepackage{braket}
\usepackage{color}      
\usepackage[pdfpagelabels,plainpages=false]{hyperref}   

\newcommand{\Fig}[1]{figure~\ref{#1}}
\newcommand{\Sec}[1]{section~\ref{#1}}

\newcommand{\bz}{\bi{z}}
\newcommand{\edzz}{\braket{\delta(z,z')}}
\newcommand{\ezs}{\braket{\tilde{z}}}
\newcommand{\ez}{\braket{z}}
\newcommand{\norm}[1]{\|#1\|}

\newcommand{\skippart}[1]{}
\newcommand{\eqref}[1]{(\ref{#1})}

\DeclareMathOperator{\tTr}{tTr}
\def\figpath{.}

\begin{document}

\title{Symmetry Breaking and Convex Set Phase Diagrams for the q-state Potts Model}

 \author{V~Zauner-Stauber$^{1}$ and F~Verstraete$^{1,2}$}
  \address{$^{1}$Vienna Center for Quantum Technology, University of Vienna, Boltzmanngasse 5, 1090 Wien, Austria}

 \address{$^{2}$Ghent University, Krijgslaan 281, 9000 Gent, Belgium}

\ead{valentin.stauber@univie.ac.at}

\begin{abstract}
We demonstrate that the occurrence of symmetry breaking phase transitions together with the emergence of a local order parameter in classical statistical physics is a consequence of the geometrical structure of probability space. To this end we investigate convex sets generated by expectation values of certain observables with respect to all possible probability distributions of classical q-state spins on a two-dimensional lattice, for several values of q. The extreme points of these sets are then given by thermal Gibbs states of the classical q-state Potts model. As symmetry breaking phase transitions and the emergence of associated order parameters are signaled by the appearance ruled surfaces on these sets, this implies that symmetry breaking is ultimately a consequence of the geometrical structure of probability space. In particular we identify the different features arising for continuous and first order phase transitions and show how to obtain critical exponents and susceptibilities from the geometrical shape of the surface set. Such convex sets thus also constitute a novel and very intuitive way of constructing phase diagrams for many body systems, as all thermodynamically relevant quantities can be very naturally read off from these sets.

\end{abstract}


\maketitle

\section{Introduction}
In a series of ground breaking papers in the late 19$^{\rm th}$ century, Gibbs \cite{Gibbs1873,Gibbs1876,MaxwellBook} elegantly derived the thermodynamic stable state of a given substance through the minimization of some thermodynamic potential (later known as the free energy), in fact by means of a geometric construction. In particular, Gibbs considered a surface given by the possible values of the thermodynamic extensive quantities (such as e.g. energy, volume and entropy) of a system of interest and realized that points on this surface with tangent planes of equal orientation  correspond to possible stable states of the substance at a temperature and pressure given by the orientation of the tangent plane. If two (or more) points belong to the same tangent plane, the corresponding states can coexist in equilibrium, characteristic for first order phase transitions. If two or more points have tangent planes with equal orientation but different distance to the origin, the state whose tangent plane is closer to the origin is metastable, corresponding to a supercritical system \cite{MaxwellBook}.

This geometrical construction can be interpreted as identifying the thermodynamically stable states as the extreme points of a convex set consisting of all possible realizable values of the thermodynamic extensive quantities of a given system. In the case of Gibbs' construction the relevant convex set is essentially the convex hull of the thermodynamic surface, termed ``secondary surface'' by Maxwell (who also produced a plaster clay model of the surface for water as a present to Gibbs in 1874). All thermodynamic properties of a system of interest can then be read off from the geometric features of this set and phase transitions correspond to non-analyticities on the surface, which arise by considering convex hulls of analytic functions \cite{VdW}. 

In this paper we extend previous work \cite{Convex1} and construct in full detail convex set thermodynamic surfaces for a paradigmatic model of classical statistical mechanics on a lattice, namely the Ashkin-Teller-Potts model \cite{AshkinTeller43,Potts52,Wu82,BaxterBook}. As in the case of Gibbs' original thermodynamic surface, the extensive quantities of the system are in competition with each other and stable states, which constitute the thermodynamic surface, are again those that minimize the free energy.

While Gibbs' original construction is capable of detecting regions of phase coexistence at first order phase transitions, they however show no signatures at second order phase transitions, as there the thermodynamic extensive quantities vary continuously across the critical point. In \cite{Convex1} it is demonstrated that by including the order parameter corresponding to such a phase transition as an extensive quantity into these sets, phase transitions are signaled through the appearance of characteristic geometrical features in the form of ruled surfaces. As these sets exist as a collection of all possible realizable states of a given system without any prior reference to any Hamiltonian which generates dynamics, the reason for the occurrence of symmetry breaking phase transitions thus lies in the geometry of the space of all possible realizable states.

Similar convex set pictures have been studied in the context of the $N$-representability problem in quantum chemistry \cite{N-representability1,N-representability2,N-representability3,N-representability4}, but without including order parameters.

In the following we will construct convex set thermodynamic surfaces for the q-state Potts model and study its geometrical features. In \Sec{sec:model_set} we construct and discuss these sets, in particular in \Sec{sec:symmbreaking} we demonstrate how symmetry breaking leads to characteristic ruled surfaces and flat parts, which are a signature of symmetry breaking phase transitions. In sections \ref{sec:topplane} and \ref{sec:leftplane} we describe additional features of the surface where the model at zero temperature becomes equivalent to coloring problems or hard-square lattice-gas models with nearest neighbor exclusion. We further describe in \Sec{sec:critexp} how to obtain critical exponents and susceptibilities from a given convex set surface. 
We conclude with final remarks and outlooks in \Sec{sec:conclusion}.
We additionally give information about the tensor network representations used to obtain numerical data in \ref{sec:TNrep} and show scatter plots generated by drawing random probability distributions of spin configurations in the Potts model in \ref{sec:scatter}.

\section{The Potts Model and its Convex Set Representations}
\label{sec:model_set}

The $q$-state Potts model \cite{AshkinTeller43,Potts52,Wu82,BaxterBook} is a generalization of the ubiquitous $\mathbb{Z}_{2}$-symmetric Ising model \cite{Ising25,Brush67}  to $\mathbb{Z}_{q}$-symmetry. It has been shown to be correspond to a $\mathbb{Z}_{q}$ lattice gauge theory of matter \cite{Balian75,Kogut79} and in certain parameter regimes to coloring problems \cite{BaxterIce70,BaxterHex70} and hard-square lattice-gas models with nearest neighbor exclusion (1NN) \cite{Baxter1NN99}.

The Potts model in a magnetic field is defined by the Hamiltonian
\begin{equation}
 H(\bz) = -J\sum_{\braket{ij}} \delta(z_{i},z_{j}) - h\sum_{j}z_{j},
 \label{eq:PottsHam}
\end{equation} 
where $z_{i}=1,\ldots,q$ is a $q$-state spin on site $i$, $\braket{ij}$ denotes nearest neighbors and $\delta$ is the Kronecker delta function. We consider the model in two spatial dimensions on a square lattice. At zero field, where the model possesses $\mathbb{Z}_{q}$-symmetry, it undergoes a symmetry breaking phase transition at some finite critical inverse temperature $\beta_{c}=\log(\sqrt{q}+1)$ \cite{Potts52,Hinterman78}, where for $\beta>\beta_{c}$ the $\mathbb{Z}_{q}$-symmetry is spontaneously broken. For $q=2$ the Potts model is equivalent to the classical Ising model \cite{Ising25} and can thus be solved exactly in zero field for all temperatures \cite{BaxterBook,Onsager44}. For general $q>2$ and zero field the model can be mapped onto a staggered six-vertex model, which can be solved exactly only at criticality \cite{TempLieb71,Baxter73}. Other solvable cases include $J<0$ at $T\to0$ and zero field for $q=3$ on the square lattice \cite{BaxterIce70}, and $q=4$ on the hexagonal lattice as well as $q=3$ on the Kagome lattice \cite{BaxterHex70}.

The symmetry breaking phase transition in zero field is continuous for $q\leq4$ and of first order for $q>4$ \cite{Potts52}. The nature of the phase transition will become apparent from the geometrical features of the corresponding convex set phase diagrams which we construct below. 

Consider the space of all possible probability distributions $P(\bz)$ of configurations of $q$-state spins $z_{i}=1,\ldots,q$ with $i$ the position on a two-dimensional square lattice with $N$ sites, which form a convex set in some high-dimensional parameter space. In particular we consider three-dimensional projections of this set in the thermodynamic limit $N\to\infty$, parameterized by the three observables 
\textit{nearest neighbor interaction energy per site}
\begin{equation}
\edzz = \frac{1}{2N}\sum_{\braket{ij}}\braket{\delta(z_{i},z_{j})},
\label{eq:edzz}
\end{equation}
\textit{shifted magnetization per site}
\begin{equation}
\ezs = \ez-\frac{q+1}{2} = \frac{1}{N}\sum_{j}\braket{z_{j}}-\frac{q+1}{2}
\label{eq:ez}
\end{equation} 
and \textit{entropy per site}
\begin{equation}
s=-\frac{1}{N}\sum_{\bz}P(\bz)\log(P(\bz)),
\label{eq:s}
\end{equation} 
where $\braket{\ldots}$ denotes expectation values with respect to $P(\bz)$. The convex set $\mathcal{C}$ is then given by all possible points $\bi{X}=[\edzz,\ezs,s]$, such that $\edzz$, $\ezs$ and $s$ are compatible with each other, i.e. they stem from a common valid probability distribution $P(\bz)$. This is an instance of the classical marginal problem \cite{Kly02,Marg91,Marg96,Marg97}. Notice that we are using a shifted magnetization with an offset $\frac{q+1}{2}$, such that the convex set is reflection symmetric with respect to $\ezs$. The extreme points on the surface of this set are then naturally given by Gibbs states of \eqref{eq:PottsHam}.

To see this, consider (hyper)planes in this three-dimensional parameter space, which are defined as families of points $\bi{X}\in\mathcal{C}$, related by a plane equation of the form
\begin{equation}
\bi{n}\cdot \bi{X} = n_{x}\edzz + n_{y}\ezs + n_{z}s = \norm{\bi{n}}\, d,
\label{eq:planeeq}
\end{equation} 
where $\bi{n}$ is the normal vector of the plane and $d$ is the distance of the hyperplane to the origin. Setting $n_{x}=2J$, $n_{y}=h$ and $n_{z}=T$, this yields exactly the (negative of the) \textit{free energy per site} of \eqref{eq:PottsHam}
\begin{equation}
 -f = 2J\edzz +h\ez + Ts,
 \label{eq:PottsFreeEnergy}
\end{equation}
where the factor 2 comes from the fact that every site has 4 nearest neighbors on a two-dimensional square lattice.
\footnote{On a general isotropic lattice the free energy is given by $-f = \frac{JK}{2}\edzz +h\ez + Ts$, where $K$ is the coordination number of the lattice.}. 

For a given set of parameters (i.e. normal vector) the hyperplane \textit{tangent} to the convex set has maximum possible distance from the origin and thus also minimizes the free energy, which is the definition of a Gibbs state. Every point on the surface thus corresponds to a state of thermodynamic equilibrium, at parameters given by the orientation of the tangent plane and free energy proportional to the distance of the tangent plane to the origin. Conversely, every point inside the convex set corresponds to a possible non-equilibrium state of the system.

If the tangent plane touches the convex set at a unique point only, then the thermodynamic stable state is unique and exactly given by a Gibbs state which yields the observables given by the tangent point for the parameters $(J,h,T)$ defined by the orientation of the tangent plane, i.e. its normal vector $\bi{n}$. If however the tangent plane touches the set on an entire line or even a plane, then the state which minimizes the free energy for these parameters is not unique, which is a prerequisite of symmetry breaking. The set of valid states can then be parameterized by one (or more) real parameters. Such ruled surfaces (continuous sets of tangent lines) or planes are thus the geometrical signatures that will enable us to detect symmetry breaking and the emergence of a connected order parameter.

\begin{figure}[p]
 \centering
 \includegraphics[width=\linewidth]{\figpath/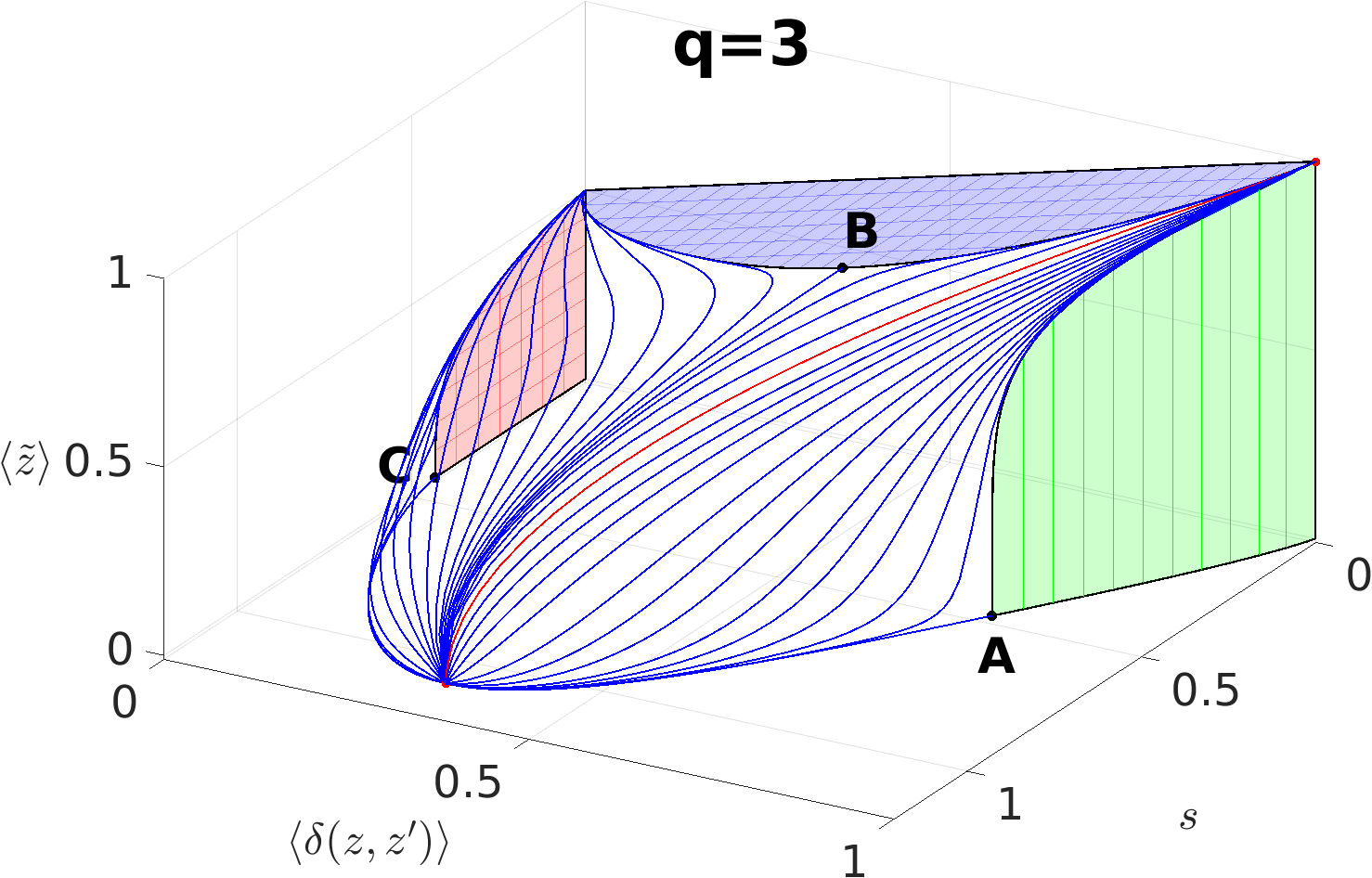}
 \caption{
 Convex set generated by nearest-neighbor interaction energy $\edzz$, shifted magnetization $\ezs$ and entropy per site $s$ of all possible probability distributions of $3$-state spins on a two-dimensional square lattice. We plot the surface of this set, corresponding to Gibbs-states of \eqref{eq:PottsHam} for $q=3$. Due to reflection symmetry we only plot the upper half of the set. 
 Blue lines denote points of constant $J=\pm 1$ and $h$ and varying temperature $T$. The red line denotes the exactly solvable decoupled case $J=0$ and thus separates regions of ferromagnetic and antiferromagnetic coupling.
 At the critical point \textbf{A} the emergence of a (green) ruled surface signals a non-uniqueness of the thermal equilibrium state at zero field and thus symmetry breaking. As a guide to the eye we have plotted a few vertical lines on the ruled surface, along which the tangent plane touches the convex set. Point \textbf{B} marks the end point of the bifurcation line of $J=-1$, $h=4$ and $T\to 0$, leading up to the (blue) top plane where the lowest energy state is exponentially degenerate, resulting in a finite residual entropy as described in \Sec{sec:topplane}. A similar situation arises at point \textbf{C}, corresponding to the end point of the line $J=-1$, $h=0$, $T\to 0$. There again the lowest energy state is exponentially degenerate, resulting in a finite residual entropy as described in \Sec{sec:leftplane}. This plane is only present for $q>2$ and does therefore not appear in the convex set drawn for the Ising model in \cite{Convex1}. As a guide to the eye we have drawn two-dimensional grids onto the top and left plane, emphasizing the fact that there the tangent plane touches the set on the entire respective planes.
 }
 \label{fig:Convex_q3}
\end{figure}

\begin{figure}[p]
 \centering
 \includegraphics[width=\linewidth]{\figpath/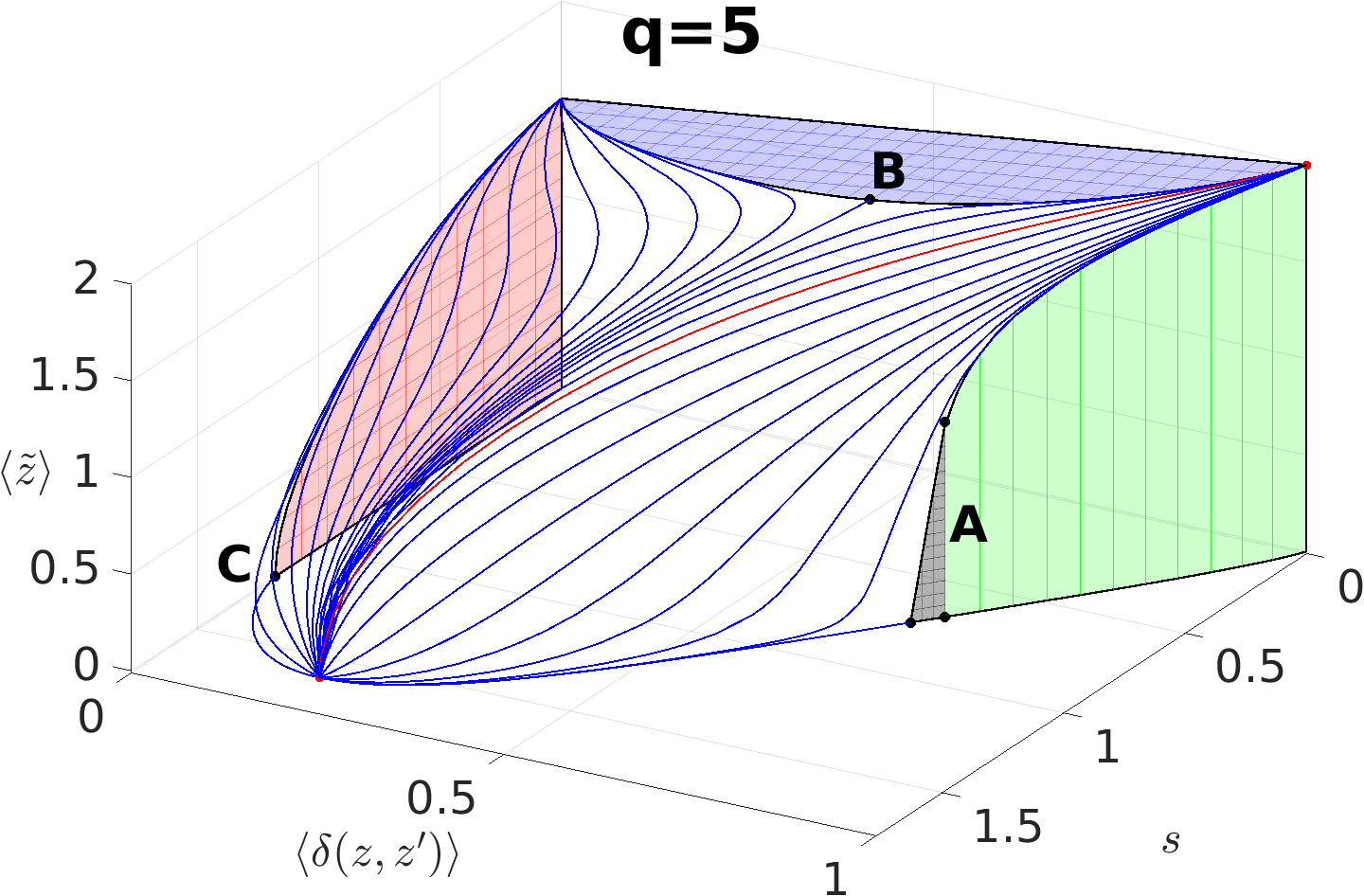}
 \caption{
 Convex set generated by the same observables as in \fref{fig:Convex_q3} for the case of $5$-state spins where the surface of this set is given by Gibbs-states of \eqref{eq:PottsHam} for $q=5$. 
For $q>4$ the phase transition is of first order and thus comes with a discontinuity of the three observables at the critical point. This results in a coexistence region of the ordered and disordered phases and the critical point \textbf{A} gets stretched out into a (gray) flat triangular surface, where any mixture of the two phases is a valid state, i.e. the two phases coexist. This flat part then smoothly connects to the (symmetry broken) ordered phase represented by the green ruled surface. As a guide to the eye we have drawn a two-dimensional grid onto the flat triangular surface, emphasizing the fact that there the tangent plane touches the set on the entire triangular surface and we have also plotted a few vertical lines on the green ruled surface, along which the tangent plane touches the convex set.
The flat surfaces emerging from points \textbf{B} and \textbf{C} are the same as described in \Fig{fig:Convex_q3}.
}
 \label{fig:Convex_q5}
\end{figure}

We show  the surfaces of the resulting convex sets for the Potts model for $q=3$ and $q=5$ in figures \ref{fig:Convex_q3} and \ref{fig:Convex_q5} respectively (for the special case of the Ising model, corresponding to $q=2$, see \cite{Convex1}). These sets show interesting geometrical features from which a wealth of other information, such as the nature of phase transitions, locations of critical points, critical exponents, susceptibilities, etc. can be extracted. The numerical data for plotting these surfaces has been obtained by means of tensor network techniques described in \ref{sec:TNrep}. For scatter plots of points obtained from random probability distributions, which approximate the convex set from the inside, see \ref{sec:scatter}.

\clearpage

\subsection{Symmetry Breaking and the Ruled Surface}
\label{sec:symmbreaking}

For zero field, $J>0$ and $T<T_{c}$ the thermodynamic state that minimizes the free energy is $q$-fold degenerate and the $\mathbb{Z}_{q}$-symmetry can be spontaneously broken, such that $\ezs\neq0$. The maximum possible value $\ezs_{\rm max}$ can then be taken as the order parameter associated to this phase transition 
\footnote{Given $\ezs_{\rm max}$ the shifted magnetization is then $\ezs=(k-(q+1)/2)\ezs_{\rm max}$ with $k=1,\ldots,q$ the integer enumerating the maximally symmetry broken states, characterized by one-site marginal distributions given by $p(z)=1/q+p(2\delta_{z,k}-1)$, where $p<1/q$ is a function of $T$. Other order parameters for the Potts model have also been proposed. One possibility for defining an observable whose expectation value in the symmetry broken phase is independent of $k$ is e.g. given by defining $O(z)=\exp(2\pi\rmi z/q)$ and measuring $\left|\braket{O}\right|=pq\in[0,1]$. 
}
. For a given set of parameters any state within this $q$-fold degenerate space thus minimizes the free energy and is characterized by the same values for $\edzz$ and $s$, but different $\ezs$ \footnote{Mixtures of maximally symmetry broken states generally do not correspond to physically realizable states as they cannot be converted into each other by means of local modifications. Mathematically they are elements of disjoint Hilbert space sectors \cite{Strocchi,Fannes}. A hint towards this fact is given by the peculiar structure of the random scatter plots for quantum and classical systems is shown in \Sec{sec:scatter}.}

This is nicely reflected in the convex sets through the emergence of a (green) ruled surface at the critical point. Zero field implies tangent planes with normal vectors lying in the $\ezs=0$ plane, i.e. $\bi{n}=[2J,0,T]$. The tangent plane touches the convex set on a unique point in the $\ezs=0$ plane everywhere except for $J>0$ and $T<T_{c}$, where the tangent plane in fact touches the convex set along a whole line for each $J$ and $T$, given by $\bi{X}(t)=\left[\edzz,t\ezs_{\rm max},s\right]$ with $t\in[-1,1]$ and $\ezs_{\rm max}>0$ the maximum value of the order parameter. An infinitesimal value of $h\neq0$ then immediately explicitly breaks the symmetry and causes the tangent plane to touch the set on a \textit{unique} point of the set infinitesimally close to the edge of the ruled surface. Or equivalently, the curve of tangent points of a tangent plane given by $\bi{n}=[2J,h\neq0,T]$ as $h\to0^{\pm}$ will end in a point with $\ezs=\pm\ezs_{\rm max}\neq 0$ for $T<T_{c}$. This nicely reflects the fact that the order parameter can be obtained by first taking the thermodynamic limit at nonzero field before letting the field go to zero.

The nature of the phase transition changes from continuous to first order for $q>4$, where a first order phase transition is characterized by a latent heat and a discontinuity of first derivatives of the free energy at the critical point. The internal energy and all other expectation values that can be written as a derivative of the free energy, such as the order parameter and also the entropy per site $s$ therefore have a discontinuity at the critical point. In the convex set we can thus detect first order phase transitions through the appearance of \textit{flat hyperplanes} at the boundary that arise \textit{even without additionally plotting the order parameter}. 
At the critical point the thermal equilibrium state is not unique and any point on this hyperplane is a valid state of the system \textit{at the critical temperature}. This corresponds to the coexistence of phases at the critical point which is characteristic for first order phase transitions. In the case of the Potts model, this flat hyperplane then smoothly connects to the ruled surface representing the symmetry broken phase $T<T_{c}$ (c.f. \Fig{fig:Convex_q5}).

\begin{figure}[t]
 \centering
 \includegraphics[width=\linewidth,keepaspectratio=true]{\figpath/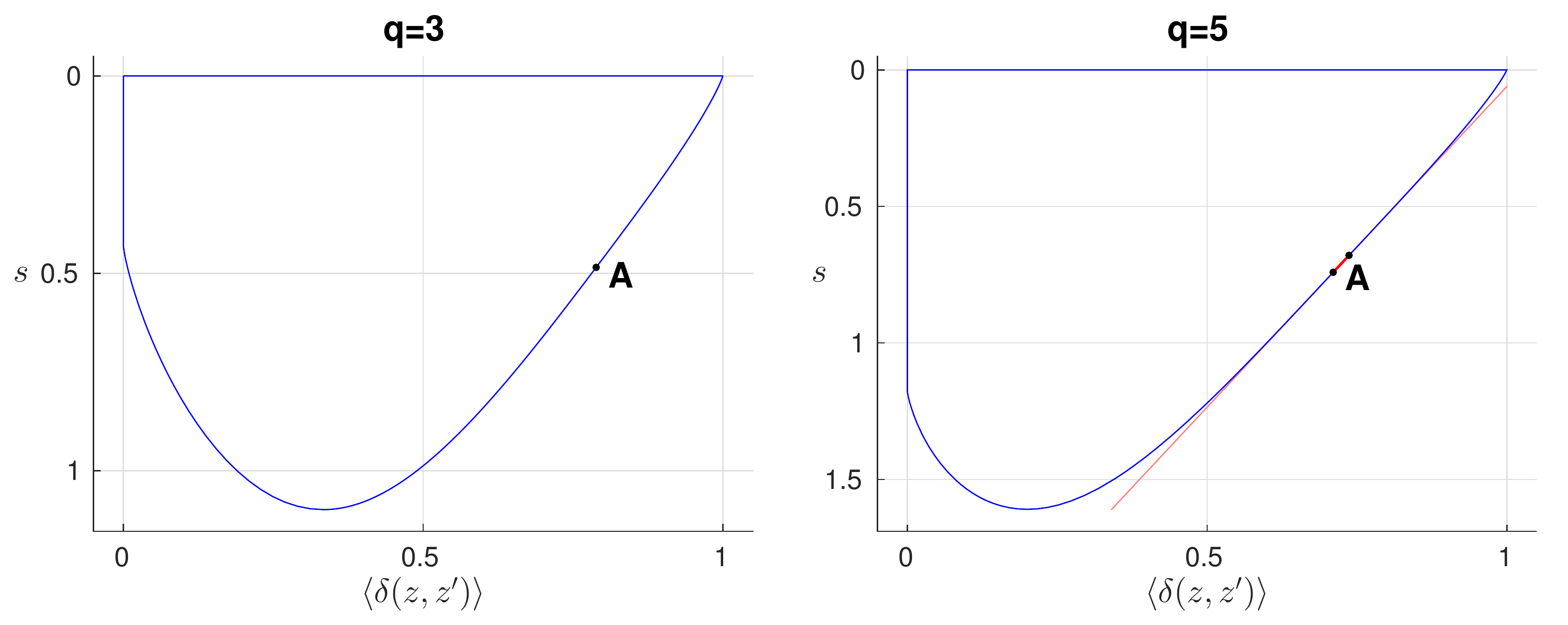}
 \caption{Surfaces of the \textit{two-dimensional} convex sets generated by nearest neighbor interaction $\edzz$ and entropy per site $s$ for the zero field Potts model for $q=3$ and $q=5$. The phase transition is continuous for $q=3$ and cannot be detected from the convex set without adding an additional axis corresponding to the order parameter $\ezs$. For $q=5$ the phase transition is however of first order and can thus be detected through the discontinuities of $\edzz$ and $s$ across the critical point \textbf{A}, which gets stretched into a straight (red) line where the two phases can coexist. As a guide to the eye we have extended this line to both sides to see that there is (albeit very small) curvature to both sides of the phase coexistence part.
 }
 \label{fig:2dsets}
\end{figure}

For continuous phase transitions the thermodynamic state at the critical point is still unique and there is no such additional hyperplane. We can thus already detect first order phase transitions in the lower dimensional convex set that does not include the order parameter. In the case of the Potts model, a two-dimensional convex set parameterized by $\edzz$ and $s$ thus already suffices to detect the phase transition for $q>4$, it will however show no signature of the phase transition for $q\leq 4$ (see \Fig{fig:2dsets}), for which adding an additional axis corresponding to the order parameter $\ezs$ is necessary.

We want to emphasize here that these convex sets and thus also the ruled surfaces exist prior to making any references to any model Hamiltonian, we just consider finite dimensional projections of the convex set of all possible probability distributions of a system of physical degrees of freedom. This means that the reason for the occurrence of symmetry breaking phase transitions ultimately lies in the geometrical structure of the space of all possible probability distributions. It would therefore be interesting to investigate all possible projections of this set and classify all possible ruled surfaces that can arise on such projections.

\subsection{Top Plane}
\label{sec:topplane}

\begin{figure}[ht]
 \centering
 \includegraphics[width=\linewidth,keepaspectratio=true]{\figpath/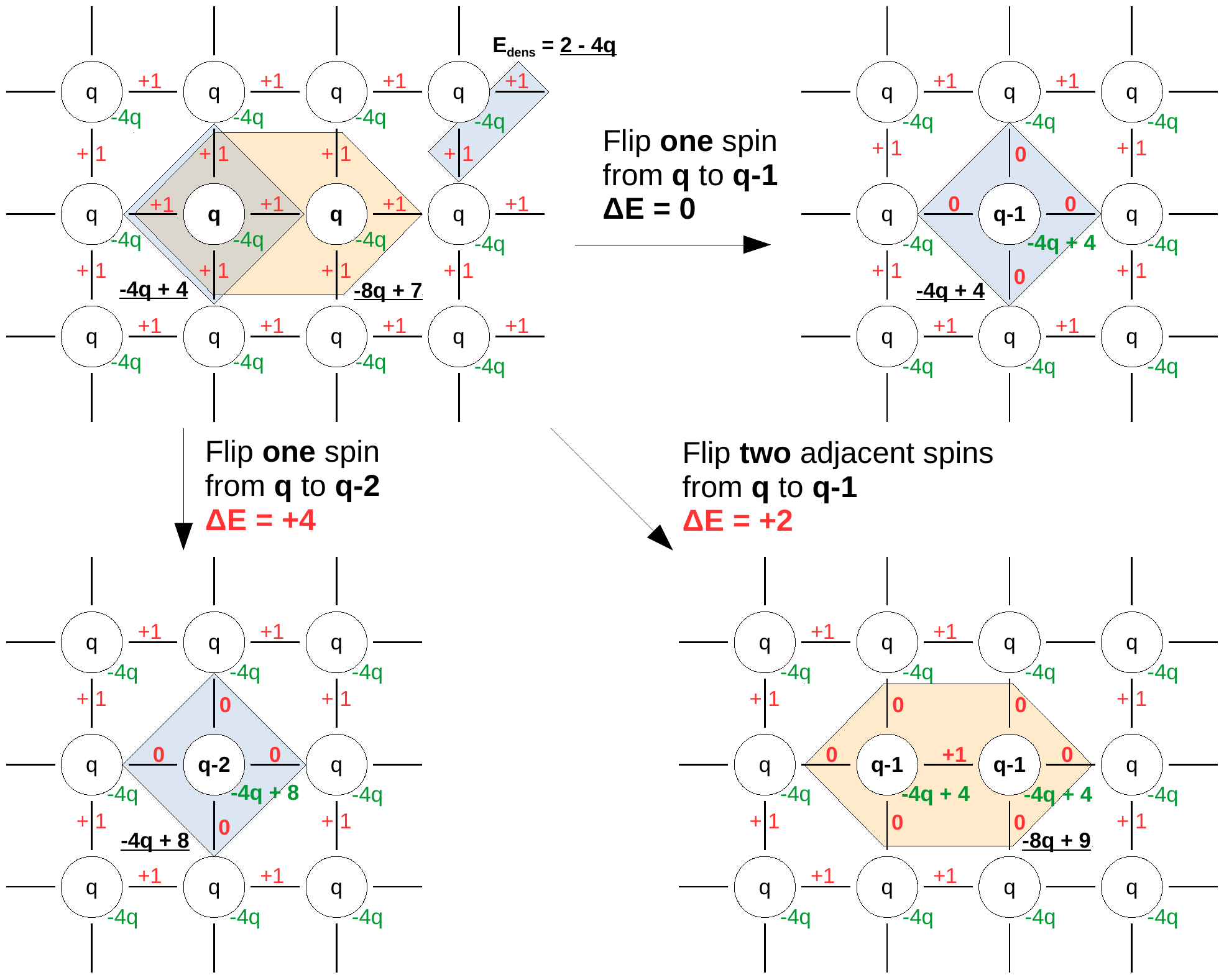}
 \caption{Construction of the degenerate space of lowest energy configurations for the top plane, corresponding to $J=-1$, $h=4$ and $T=0$. Starting from the fully polarized state $z_{i}=q$ with lowest possible energy, flipping single spins from $q$ to $q-1$ leaves the overall energy invariant. Flipping two or more adjacent spins however results in a net energy increase, as does flipping from $q$ to any $z<q-1$. The resulting space consists of all configurations where $z_{j}\in[q,q-1]$ such that every $z_{i}=q-1$ is completely surrounded by $z_{j}=q$.}
 \label{fig:top_plane}
\end{figure}

The top (blue) plane corresponds to parameters $J=-1$, $h=4$ and $T=0$ where the tangent plane touches the convex set on the entire top plane, meaning that the thermal equilibrium state is not unique and in fact all states on this plane are valid equilibrium states for these parameters.

Exactly at this point, the two terms in the Hamiltonian become ``equally strong'' in the following sense. If we start from the completely polarized state $z_{j}=q$, the magnetic field term is minimized, whereas the interaction part has a positive energy contribution, resulting in a net energy of $2-4q$ per site. If we now flip one spin at an arbitrary position from $q$ to $q-1$, we gain exactly the same amount of energy from the interaction term as we lose from the magnetic field term and the overall energy stays the same. We can now continue flipping spins that way without changing the energy, as long as we never flip any spins \textit{next to an already flipped one}, which would result in a net energy increase of $+2$. In general, a cluster of $N_{f}$ flipped spins and a boundary of length $N_{b}$  results in a net energy change of $4N_{f}-N_{b}\geq0$, which is only zero for $N_{f}=1$. The two lowest energy states with the smallest magnetization are thus the two N\'eel states between $q$ and $q-1$. Similarly, flipping from $q$ to any $z<q-1$ always results in a net energy increase and the restricted space of lowest energy states is thus given by all configurations $z_{j}\in[q,q-1]$ such that every $z_{i}=q-1$ is completely surrounded by $z_{j}=q$ (see also \Fig{fig:top_plane}). This restricted space is equivalent to the configuration space for the nearest-neighbor exclusion lattice-gas model (1NN) \cite{Baxter1NN99} and grows exponentially with the system size. 

At $T=0$ all such configurations are equally likely; the entropy per site is therefore finite and measures the exponential growth of the space of lowest energy configurations. This symmetry of equal probability can however be spontaneously broken as any statistical mixture of such configurations is a valid state of the system with equal free energy $f=2-4q$. The entirety of all such mixtures is exactly given by the top blue plane in the convex sets, where point \textbf{B} marks the state of equal probability which has maximal entropy.

To calculate the boundary of the top blue plane we consider tiny perturbations away from this point in parameter space, which immediately cause a jump onto the edge of the plane. Similar to degenerate perturbation theory we then simulate this perturbation Hamiltonian only within the restricted subspace of the top plane to lift the exponential degeneracy and determine its extreme points. The perturbation Hamiltonian is just the magnetic field term
\begin{equation}
 \beta H_{1} = \mu\sum_{j}z_{j},
\end{equation} 
where $\mu$ is usually small. Since we however simulate this Hamiltonian in the \textit{restricted} subspace only (which also makes the simulation non-trivial), $\mu$ need not be small and just controls the position along the edge of the top plane. We therefore wish to evaluate
\begin{equation}
 Z=\sum_{\bz\in\mathcal{Z}_{t}} \rme^{-\mu\sum_{j}z_{j}}
 \label{eq:topedge}
\end{equation} 
where the sum is only over the space of valid configurations $\mathcal{Z}_{t}$ given by the top plane and $\mu\in\mathbb{R}$. The entropy per site $s$ is then given by
\begin{equation}
 s = \log(z)+\mu\ez
 \label{eq:splane}
\end{equation} 
where $z=Z^{1/N}$ is the partition function per site. The other observables $\edzz$ and $\ezs$ are computed as usual but with respect to \eqref{eq:topedge}. Note that entropy and $\edzz$ are independent of $q$ and $\ezs$ for different $q$ are related by just an offset. The top plane thus has the  \textit{same shape for all q}, but different vertical offset in $\ezs$.

Note that \eqref{eq:topedge} is equivalent to the 1NN model in a chemical potential $\mu$ \cite{Gaunt1NN65, Guo1NN02, Fernandes1NN07}, where states $q$ and $q-1$ correspond to an empty and occupied site respectively. 
The limits $\mu\to\pm\infty$ correspond to the the completely polarized and the N\'eel states respectively (or equivalently the completely empty and maximally filled lattice respectively in terms of the 1NN model) and thus have zero entropy, while $\mu=0$ corresponds to point \textbf{B} with maximal residual entropy $s_{\rm res}$. Our calculated value at this point reproduces the (log of the) value $\kappa(1)$ given in section 1.1 of \cite{Baxter1NN99} up to machine precision. The tensor network we use to simulate \eqref{eq:topedge} is described in \ref{sec:TNtop}.

\subsection{Left Plane}
\label{sec:leftplane}

\begin{figure}[t]
 \centering
 \includegraphics[width=0.8\linewidth,keepaspectratio=true]{\figpath/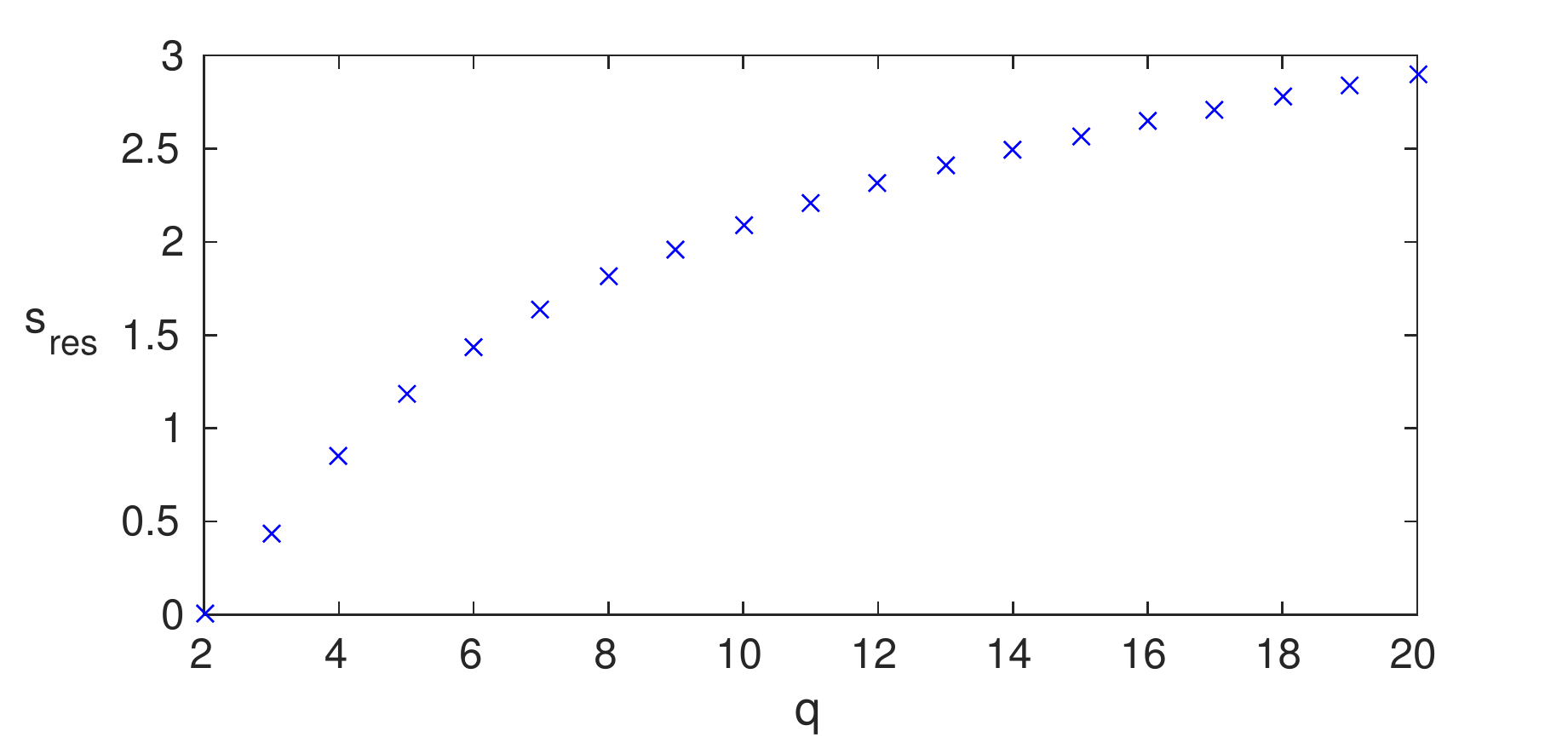}
 \caption{Residual entropy at point \textbf{C} on the left plane, given by $J=-1$, $h=0$ and $T=0$, for values $q\in[2,20]$. This corresponds to the (log of the) number of proper vertex colorings per site of a square lattice with $q$ colors.}
 \label{fig:sres_v_q}
\end{figure}

The left (red) plane with $\edzz=0$ corresponds to parameters $J=-1$, $h=0$ and $T=0$ where the tangent plane touches the convex set on the entire left plane, meaning that the thermal equilibrium state is not unique and in fact all states on this plane are valid equilibrium states for these parameters. 

For these parameters the lowest energy states are given by all configurations $z_{j}\in[1,q]$, such that \textit{no nearest neighbors are in the same state}. This is the famous vertex coloring problem and consequently, the partition function can be written as a chromatic polynomial in $q$ \cite{Fortuin72,Tutte67} and counts the number of proper vertex colorings of the two-dimensional square lattice with $q$ colors. For any $q>2$ the number of valid configurations is exponentially large in system size and we are thus presented with the same situation as for the top plane in the previous section, but with a different restricted subspace. For $q=2$ (i.e. the Ising model) this problem is trivial as only two valid configurations exist (the two N\'eel states) and the left plane is absent).

Again, at $T=0$ all these configurations are equally likely, leading to a residual, non-zero entropy $s_{\rm res}$. For $q=3$ this can be mapped onto the problem of residual entropy of square ice \cite{BaxterIce70}, for which the value is known exactly as $s_{\rm res}=3/2\log(4/3)\approx 0.431523$ \cite{LiebIce1,LiebIce2}. For $q>3$ there are no exact solutions for the square lattice. The symmetry of equal probability can again be spontaneously broken and any point on the left flat surface then corresponds to a valid statistical mixture of configurations within the restricted subspace, giving the same free energy $f=0$. All these mixtures are represented by the left red plane in the convex sets, where point \textbf{C} corresponds to the equal probability mixture which has maximal entropy $s_{\rm res}$.

To determine the boundary of the left red plane we proceed the same way as in \Sec{sec:topplane} and simulate
\begin{equation}
 Z=\sum_{\bz\in\mathcal{Z}_{c}} \rme^{-\mu\sum_{j}z_{j}}
 \label{eq:leftedge}
\end{equation}
where the sum is now over all proper vertex colorings $\mathcal{Z}_{c}$. The entropy is again given by \eqref{eq:splane}.

We have calculated $s_{\rm res}$ for several values of $q$ (see \Fig{fig:sres_v_q}), where we can reproduce the exact value for $q=3$ up to an accuracy of $\mathcal{O}(10^{-10})$ with bond dimension $D=800$ of the MPS-representation of the dominant eigenvector of the transfer matrix.

The tensor network used to simulate \eqref{eq:leftedge} is described in \ref{sec:TNleft}.

\subsection{Critical Exponents and Susceptibilities}
\label{sec:critexp}
If we are given the entire convex set as a function of the extensive observables we can determine critical exponents and susceptibilities purely from the geometrical shape of its surface, i.e. completely independent from the intensive parameters $J$, $T$ and $h$. To ease notation in this section we will write
\begin{equation}
 t:=\edzz,\qquad z:=\ezs.
\end{equation} 

Critical exponents for $q\leq4$ can be extracted from the change of the tangent plane orientation around the critical point. For this we need the functional relation between an observable and a model parameter close to the critical point. As an example consider the shifted magnetization $z$ for zero field slightly below the critical temperature $T_{c}$. 
There we expect $z$ to behave as
\begin{equation}
 z\propto\left( 1-\frac{T}{T_{c}} \right)^{b},\quad h=0
 \label{eq:beta_exponent}
\end{equation} 
with $b$ the critical exponent for the magnetization. 

We assume the thermodynamic surface to be given e.g. by the interaction energy $t$ as a function of the (independent) variables entropy $s$ and shifted magnetization $z$, i.e. $t=t(s,z)$. Our intention is to extract $b$ entirely from the geometrical form of the thermodynamic surface, i.e. from the surface given by the function $t(s,z)$. We therefore need a way to express the model parameters $J$, $T$ and $h$ in terms of the observables $t$, $s$ and $z$. From \eqref{eq:planeeq} and \eqref{eq:PottsFreeEnergy} we saw that they are precisely the elements of the normal vector to the surface function $t(s,z)$. On the other hand, the normal vector to the thermodynamic surface $t(s,z)$ at a given point is 
\begin{equation}
 \bi{n}=\left[1,-\frac{\partial t}{\partial s},-\frac{\partial t}{\partial z}\right].
 \label{eq:normalvec}
\end{equation} 
and we can immediately identify
\begin{equation}
 T=-2J\,\frac{\partial t}{\partial s},\qquad h=-2J\,\frac{\partial t}{\partial z}
 \label{eq:params_from_normalvec}
\end{equation} 

Without loss of generality we fix $J=1$ and consider the case $h=0$, i.e. the path of normal vectors with $n_{3}=-\frac{\partial t}{\partial z}=0$. We can then write
\begin{equation}
 \log z=b\log\left[ 1- \frac{\partial t}{\partial s}\left( \frac{\partial t}{\partial s}\Big|_{\bi{A}} \right)^{-1} \right] + {\rm const}
 \label{eq:beta_exponent_convex}
\end{equation} 
where we have extracted the critical temperature from the orientation of the tangent plane at the critical point $\bi{A}$ as $T_{c}=-2\,\frac{\partial t}{\partial s}\big|_{\bi{A}}$. If we plot $\log\Big[ 1- \frac{\partial t}{\partial s}\left( \frac{\partial t}{\partial s}\Big|_{\bi{A}} \right)^{-1} \Big]$ vs. $\log z$ we expect a linear relation near $\bi{A}$ and we can read off $b$ from the slope \footnote{As per definition of the ruled surface, $z$ is not unique along this path and it is understood that we take the maximum of $z$ in \eqref{eq:beta_exponent_convex} for each $s$ and $t$, i.e. the order parameter. This path is nothing but the upper boundary of the ruled surface shown e.g. in \fref{fig:Convex_q3}. Alternatively we could have formulated \eqref{eq:beta_exponent_convex} in terms of derivatives of $s=s(t,z)$. Notice however that $z=z(s,t)$ is not a good choice as it is a highly multivalued function on the ruled surface.}.


Estimates for the critical exponents calculated that way from the obtained given numerical data are of the same accuracy as estimates obtained from conventional fits of observables vs. model parameters (i.e. a logarithmic fit of \eqref{eq:beta_exponent}).

Furthermore, susceptibilities defined as the derivatives of the (extensive) observables $t$, $s$ and $z$ with respect to the (intensive) model parameters $J$, $T$ and $h$ can be calculated from the curvature of the surface. Loosely speaking, we would like to know how we move on the surface if we change the orientation of the normal vector infinitesimally along one component. In other words, if we change the orientation of $\bi{n}$ by $\delta T$ along $n_{2}$, what are the resulting $\delta t$, $\delta s$ and $\delta z$. The relation between these changes is of course dictated by the function $t(s,z)$ (or in fact any other representation of the surface, e.g. as $s(t,z)$ or $z(t,s)$).

With fixed $J=1$ we have established the model parameters as functions purely of the observables in \eqref{eq:params_from_normalvec}, i.e. we consider the vector-valued function
\begin{equation}
 \bi{p}(s,z) = [T(s,z),h(s,z)].
\end{equation} 
According to \eqref{eq:normalvec} the Jacobian of this function is then proportional to the Hessian of $t(s,z)$ via
\begin{equation}
 J_{\bi{p}}=
\left[\begin{array}{cc} 
\frac{\partial T}{\partial s} & \frac{\partial T}{\partial z}\\
\frac{\partial h}{\partial s} & \frac{\partial h}{\partial z}
\end{array}\right] 
=-2
\left[\begin{array}{cc} 
\frac{\partial^{2} t}{\partial s^{2}} & \frac{\partial^{2} t}{\partial s \partial z} \\
\frac{\partial^{2} t}{\partial s \partial z} & \frac{\partial^{2} t}{\partial z^{2}} 
\end{array}\right] ,
\label{eq:jacobian}
\end{equation} 
so we can express it purely in terms of the observables. The infinitesimal change in the normal vector when moving infinitesimally on the surface is then given by $\delta\bi{p} = J_{\bi{p}}\cdot \delta\bi{O}$

We are however interested in the converse direction, i.e. the derivatives which are the elements of the Jacobian of the inverse function $\bi{O}(T,h):=\bi{p}^{-1}(T,h)=[s(T,h),z(T,h)]$
\begin{equation}
  J_{\bi{O}}=
\left[\begin{array}{cc} 
\frac{\partial s}{\partial T} & \frac{\partial s}{\partial h}\\
\frac{\partial z}{\partial T} & \frac{\partial z}{\partial h}
\end{array}\right] .
\end{equation} 
The inverse function theorem then gives the elements of this Jacobian by inverting \eqref{eq:jacobian} and we can thus obtain the susceptibilities from the second derivatives of $t(s,z)$, i.e. we can obtain $\delta\bi{O} = J_{\bi{O}}\cdot \delta\bi{p}=J_{\bi{p}}^{-1}\cdot \delta\bi{p}$. With the determinant of \eqref{eq:jacobian} given by
\begin{equation}
 \det J_{\bi{p}} = 4 \left[ \frac{\partial^{2} t}{\partial s^{2}} \frac{\partial^{2} t}{\partial z^{2}} - \left( \frac{\partial^{2} t}{\partial s \partial z} \right)^{2} \right]
 \label{eq:jacobian_det}
\end{equation}
we get e.g.
\begin{eqnarray}
 \chi_{T}  = \frac{\partial z}{\partial T}& = \frac{1}{2}\left[ \frac{\partial^{2} t}{\partial s^{2}} \frac{\partial^{2} t}{\partial z^{2}} - \left( \frac{\partial^{2} t}{\partial s \partial z} \right)^{2} \right]^{-1} \frac{\partial^{2} t}{\partial s \partial z}\\
 &=\frac{1}{2}\left[ \frac{\partial^{2} t}{\partial s^{2}} \frac{\partial^{2} t}{\partial z^{2}}\left( \frac{\partial^{2} t}{\partial s \partial z} \right)^{-1} -\frac{\partial^{2} t}{\partial s \partial z} \right]^{-1} ,\\
  \chi_{h} = \frac{\partial z}{\partial h} & = -\frac{1}{2}\left[ \frac{\partial^{2} t}{\partial s^{2}} \frac{\partial^{2} t}{\partial z^{2}} - \left( \frac{\partial^{2} t}{\partial s \partial z} \right)^{2} \right]^{-1} \frac{\partial^{2} t}{\partial s^{2}}\\
  & = \frac{1}{2}\left[ \left( \frac{\partial^{2} t}{\partial s \partial z} \right)^{2} \left(  \frac{\partial^{2} t}{\partial s^{2}} \right)^{-1} - \frac{\partial^{2} t}{\partial z^{2}} \right]^{-1}.
\end{eqnarray}
These relations are only valid if \eqref{eq:jacobian} is invertible and the susceptibilities can diverge if \eqref{eq:jacobian_det} becomes zero. 

Consider for example the magnetic susceptibility $\chi_{T}$. In \fref{fig:Convex_q3} along the path $h=0$ we have $z=0$ and $\frac{\partial t}{\partial z}=0$ (i.e. normal vectors with $n_{3}=0$), as $t$ becomes maximal when $z=0$. This is the case for any $s$ along this path, the mixed derivative $\frac{\partial^{2}t}{\partial s \partial z}$ is thus also zero everywhere. For $T>T_{c}$ the second derivative $\frac{\partial^{2}t}{\partial z^{2}}$ is finite, but becomes zero as $T\to T_{c}^{+}$ (and is in fact zero at every point on the ruled surface per definition). With $\frac{\partial^{2}t}{\partial s \partial z}=0$ and $\frac{\partial^{2}t}{\partial z^{2}}\to0$ the determinant of the Jacobian becomes zero as $T\to T_{c}$ and $\chi_{T}$ diverges.

\section{Conclusions}
\label{sec:conclusion}
We have presented an explicit construction of Gibbs' thermodynamic surface in the form of convex sets for the classical $q$-state Potts model on a two-dimensional square lattice. We established that points on these surfaces correspond to thermodynamically stable states of the model at parameters given by the orientation of the tangent plane going through that point. Points on the inside on the other hand correspond to non-equilibrium states. These convex sets also constitute a novel and very intuitive way of constructing phase diagrams for many body systems, as all thermodynamically relevant quantities are very naturally included in these sets.

In particular we have demonstrated that symmetry breaking phases appear in this sets in the form of ruled surfaces, where the thermodynamically stable state is not unique. Especially for first order phase transitions the critical point gets stretched out into a flat surface, corresponding to the coexistence of phases at the critical point, characteristic for first order phase transitions. As these sets exist in probability space of the physical degrees of freedom prior to any notion of a Hamiltonian, this implies that the occurrence of symmetry breaking phase transitions is purely a consequence of the geometrical structure of probability space. To further elucidate that point we have shown scatter plots of points obtained from random probability distributions, which all lie inside the convex set per construction and give further information about the internal structure of the constructed convex sets.

We have also identified two regimes, where the ground state at $T\to0$ is exponentially degenerate and the Potts model becomes equivalent to the vertex coloring problem and the 1NN model respectively. The corresponding flat parts in the convex sets constitute all possible states of the system in these regimes. The symmetry of equal weight superposition of these degenerate states can be spontaneously broken on and distinguished by the observables chosen to constitute the convex set, causing the emergence of these flat parts.

Additionally we have shown how thermodynamic relevant quantities such as critical exponents and susceptibilities can be extracted from the curvature of the thermodynamic surface.

In terms of projections of the set of all possible probability distributions of a physical system it remains to investigate and classify all possible ruled surfaces that can arise on such convex sets projections. Some attempts for the case of fully connected graphs have been made in \cite{Chen1,Chen2}. In the context of models of classical statistical mechanics it would be interesting to obtain equivalent convex set representations in the presence of different types of phase transitions, such as e.g. Berezinskii-Kosterlitz-Thouless phase transitions \cite{Berezinskii71, KT73}. As topological phase transitions in two-dimensional quantum many body systems appear as symmetry breaking phase transitions in the boundary theories of the entanglement degrees of freedom \cite{Shadows}, the question remains what would be the equivalent in the context of classical mechanics.
\section*{Acknowledgments}
We thank A. Gendiar, C. Dellago and M. Mari{\"e}n for inspiring discussions.
This work was supported by the Austrian Science Fund (FWF): F4104 SFB ViCoM and F4014 SFB FoQuS, ERA Chemistry and the EC through grants QUTE and SIQS.

\newpage
\clearpage
\appendix

\section{Tensor network representations for classical spin lattice models}
\label{sec:TNrep}

\begin{figure}[t]
 \centering
 \includegraphics[width=0.8\linewidth,keepaspectratio=true]{\figpath/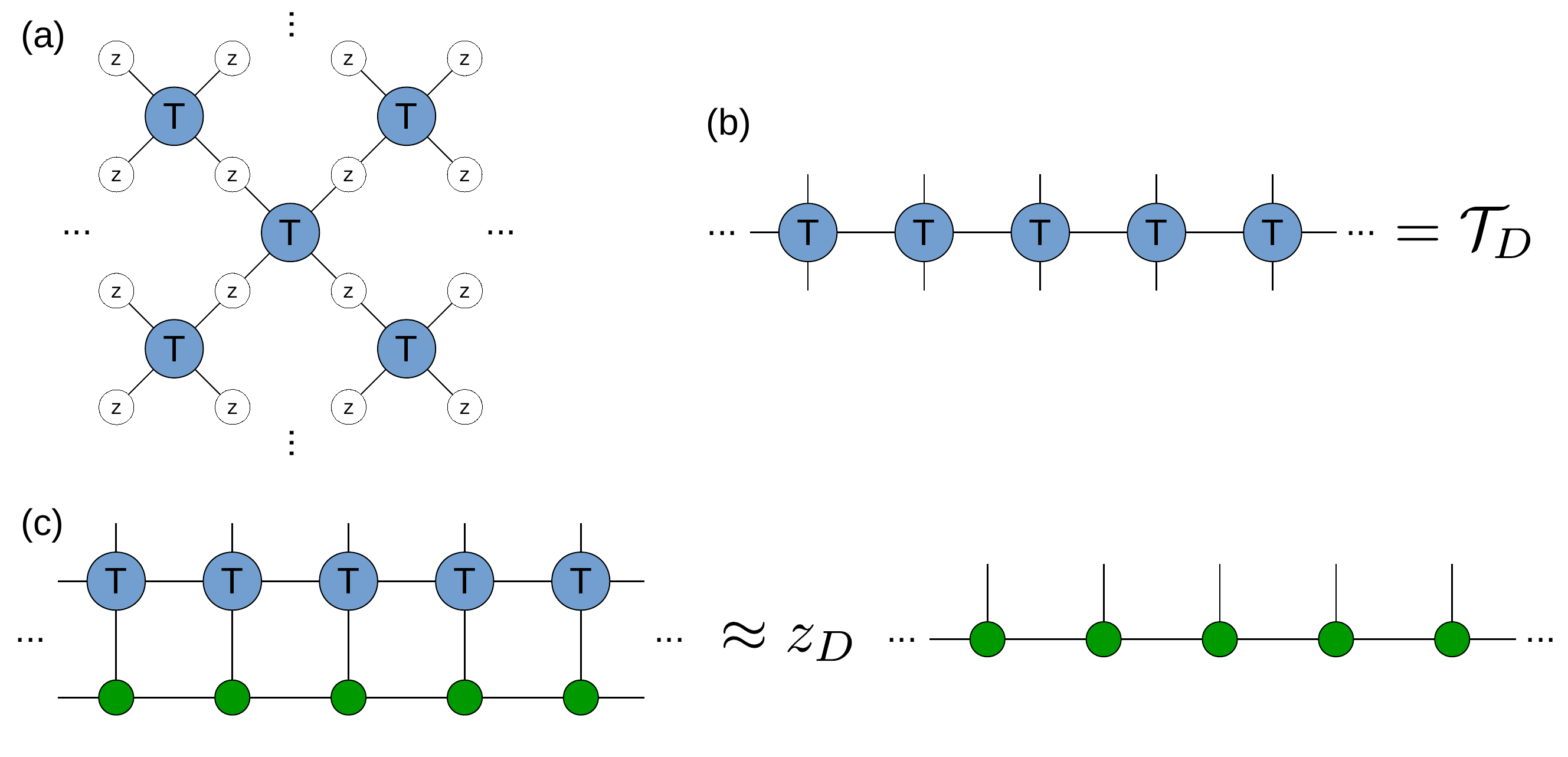}
 \caption{
 Graphical representations of the tensor networks. \textbf{(a)} Decomposition of the partition function \eqref{eq:partition} into a tensor network, specified by the MPOs $T$ given in \eqref{eq:PottsMPO}. \textbf{(b)} Concatenating MPOs along a line yields the diagonal-to-diagonal transfer matrix $\mathcal{T}_{D}$. \textbf{(c)} Approximation of the dominant eigenvector of $\mathcal{T}_{D}$ with a MPS.
 }
 \label{fig:TNreps}
\end{figure}

In this section we give information about the tensor network representations of the thermal partition function $Z$ and the tensor network methods used to approximately calculate the partition function $z$ per site, the entropy per site $s$ and the expectation values of local observables (such as $\edzz$ and $\ez$) of Gibbs states of \eqref{eq:PottsHam}.

Consider a two-dimensional square lattice with 
$N=2L^{2}$
sites where the partition function is given by 
\begin{equation}
 Z = \sum_{\bz}\exp[-\beta H(\bz)] = \sum_{\bz}\prod_{\braket{ij}}\exp\{\beta [J\delta(z_{i},z_{j})+\frac{h}{4}(z_{i}+z_{j})]\}.
 \label{eq:partition}
\end{equation} 
This can be understood as a contraction of a tensor network consisting of 4-index tensors
\begin{eqnarray}
 T_{z_{i},z_{j},z_{k},z_{l}}=& \exp\{\beta J[\delta(z_{i},z_{j}) + \delta(z_{i},z_{k}) + \delta(z_{j},z_{l}) + \delta(z_{k},z_{l})]\}\times\nonumber\\
 &\exp\left[\frac{\beta h}{2}(z_{i} +z_{j} +z_{k} +z_{l})\right].
 \label{eq:PottsMPO}
\end{eqnarray}

such that 
\begin{equation}
Z=\tTr\Big(\prod_{n=1}^{N/2}T\Big)=\sum_{\bz}T_{z_{i},z_{j},z_{k},z_{l}}T_{z_{l},z_{m},z_{n},z_{o}}T_{z_{p},z_{q},z_{j},z_{r}}T_{z_{r},z_{s},z_{m},z_{t}}\ldots,
\end{equation} 
where $\tTr$ denotes the tensor trace. Notice that every index appears exactly twice. Since every tensor \eqref{eq:PottsMPO} contains 4 nearest neighbor interaction terms and $Z$ comprises exactly $2N$ of such terms there are half as many tensors in the network as there are sites on the lattice (c.f. \fref{fig:TNreps}).

A concatenation of this choice of tensor along a line throughout the entire lattice yields the diagonal-to-diagonal transfer matrix (DTM) $\mathcal{T}_{D}$ of the partition function, or in other words, $T$ represents a Matrix Product Operator (MPO) \cite{MPO1,MPS_MPO} decomposition of the DTM. Other tensor decompositions -- e.g. yielding the row-to-row or column-to-column transfer matrix upon concatenation -- are also possible, the advantage of \eqref{eq:PottsMPO} is however that the DTM is hermitian for all $J$, $h$ and $\beta$.

We therefore have $Z=\Tr(\mathcal{T}_{D}^{L})$ and in the limit $N\to\infty$ the dominant eigenvalue of the DTM corresponds to the partition function per diagonal $z_{D}=Z^{\frac{1}{L}}$  of the system (c.f. e.g. \cite{BaxterBook}).

In order to evaluate the partition function per site $z=Z^{\frac{1}{N}}=z_{D}^{\frac{1}{2L}}$ and the local observables in the thermodynamic limit $L\to\infty$ we obtain the dominant eigenvector of the DTM by means of Matrix Product State (MPS) \cite{MPS_MPO} techniques. More specifically, we use a modification of the algorithm presented in \cite{TDVP} for MPOs in the thermodynamic limit \cite{MPO_GS} to calculate the partition function per site $z$ and an MPS approximation of the dominant eigenvector of the DTM, which can be used to calculate all local observables, in particular $\edzz$ and $\ezs=\ez-\frac{q+1}{2}$.

As we have access to the partition function per site $z$, we can now easily evaluate the entropy per site, which is given by
\begin{equation}
 s = \beta e - \beta f = \beta e + \log(z),
 \label{eq:entropy}
\end{equation} 
with the internal energy per site $e=-2J\edzz-h\braket{z}$.

\subsection{Top Plane and the 1NN}
\label{sec:TNtop}
As described in \Sec{sec:topplane} in order to determine the boundary of the top plane we simulate the trivial perturbation Hamiltonian in the restricted subspace $\mathcal{Z}_{t}$ given by all configurations $z_{j}\in[q,q-1]$ such that every $z_{i}=q-1$ is completely surrounded by $z_{j}=q$, i.e. we wish to evaluate
\begin{equation}
 Z_{\rm tp}=\sum_{\bz\in\mathcal{Z}_{t}} \rme^{-\mu\sum_{j}z_{j}}
 \label{eq:topedge_2}
\end{equation} 
There it is also mentioned that \eqref{eq:topedge_2} is equivalent to the 1NN model in a chemical potential \cite{Baxter1NN99, Gaunt1NN65, Guo1NN02, Fernandes1NN07} by interpreting $z_{j}=q,q-1$ as empty and occupied sites of a lattice gas with nearest neighbor exclusion respectively. We can therefore arrive at a formulation of \eqref{eq:topedge_2} where the entropy per site $s$ and the interaction $\edzz$ are independent of $q$ and $\ezs$ for different $q$ are related by an offset.

By substituting $z_{j}=q-s_{j}$ with $s_{j}=0,1$ we get
\begin{equation}
 Z_{\rm tp} = \sum_{\bi{s}\in\mathcal{S}}\rme^{-\mu\sum_{j}(q-s_{j})}=\rme^{-\mu q N}\sum_{\bi{s}\in\mathcal{S}}\rme^{\mu\sum_{j}s_{j}}=\rme^{-\mu q N}Z_{\rm hs},
 \label{eq:HS}
\end{equation} 
where $Z_{\rm hs}$ is the partition function of the 1NN model and $\mathcal{S}$ is the restricted set of all configurations $s_{j}\in[0,1]$ such that every $s_{i}=1$ is completely surrounded by $s_{j}=0$. The partition functions per site are then related by $z_{tp}=\rme^{-\mu q}z_{hs}$.

To evaluate $Z_{\rm hs}$ we can achieve a summation over the restricted subspace only by summing over all configurations $s_{j}\in[0,1]$, but giving configurations with neighboring $s_{i}=s_{j}=1$ statistical weight zero. This way we obtain a MPO decomposition with bond dimension 2, with MPOs given by
\begin{equation}
 T^{\rm hs}_{s_{i},s_{j},s_{k},s_{l}} = f_{s_{i},s_{j}}f_{s_{i},s_{k}}f_{s_{j},s_{l}}f_{s_{k},s_{l}}\exp\left[\frac{\mu}{2}(s_{i}+s_{j}+s_{k}+s_{l})\right],
\end{equation} 
where the $2\times 2$ matrix $f$ is given by
\begin{equation}
 f_{s_{i},s_{j}}=1-s_{i}s_{j}.
\end{equation} 

The magnetization then becomes
\begin{equation}
 \braket{z}=-\frac{\partial \log z_{\rm tp}}{\partial \mu} = q - \frac{\partial \log z_{\rm hs}}{\partial \mu} = q - \braket{s}.
 \label{eq:sexp}
\end{equation}
giving for the entropy per site
\begin{equation}
 s = \log z_{\rm tp} + \mu\braket{z} = \log z_{\rm hs} - \mu\braket{s}.
\end{equation} 

$\edzz$ is invariant as $\delta(z_{i},z_{j}) = \delta(s_{i},s_{j})$ and the expectation value is evaluated with respect to the same probability distribution.
\subsection{Left Plane and the Coloring Problem}
\label{sec:TNleft}
As described in \Sec{sec:leftplane} to determine the boundaries of the left (red) plane we seek to simulate the same perturbation Hamiltonian as in the last section
\begin{equation}
 Z_{\rm lp}=\sum_{\bz\in\mathcal{Z}_{c}} \rme^{-\mu\sum_{j}z_{j}},
 \label{eq:leftedge_2}
\end{equation}
but with $\mathcal{Z}_{c}$ a different restricted subspace, given by all configurations $z_{j}\in[1,q]$ such that no nearest neighbors have the same value.

We can again achieve a summation over the restricted subspace only by summing over all configurations $z_{j}\in[1,q]$, but giving configurations with neighboring $z_{i}=z_{j}$ statistical weight zero. This way we obtain a MPO decomposition with (unchanged) bond dimension $q$, with MPOs given by
\begin{equation}
 T^{\rm lp}_{z_{i},z_{j},z_{k},z_{l}} = f_{z_{i},z_{j}}f_{z_{i},z_{k}}f_{z_{j},z_{l}}f_{z_{k},z_{l}}\exp\left[-\frac{\mu}{2}(z_{i}+z_{j}+z_{k}+z_{l})\right],
\end{equation} 
where the $q\times q$ matrix $f$ is given by
\begin{equation}
 f_{z_{i},z_{j}}=1-\delta(z_{i},z_{j}).
\end{equation} 

The expectation value of the interaction $\edzz$ is zero per construction and the entropy is then given by
\begin{equation}
 s = \log z_{\rm lp} + \mu\braket{z}.
\end{equation}

\section{Random Scatter Plots}
\label{sec:scatter}
\begin{figure}[p]
 \centering
 \includegraphics[width=0.8\linewidth,keepaspectratio=true]{\figpath/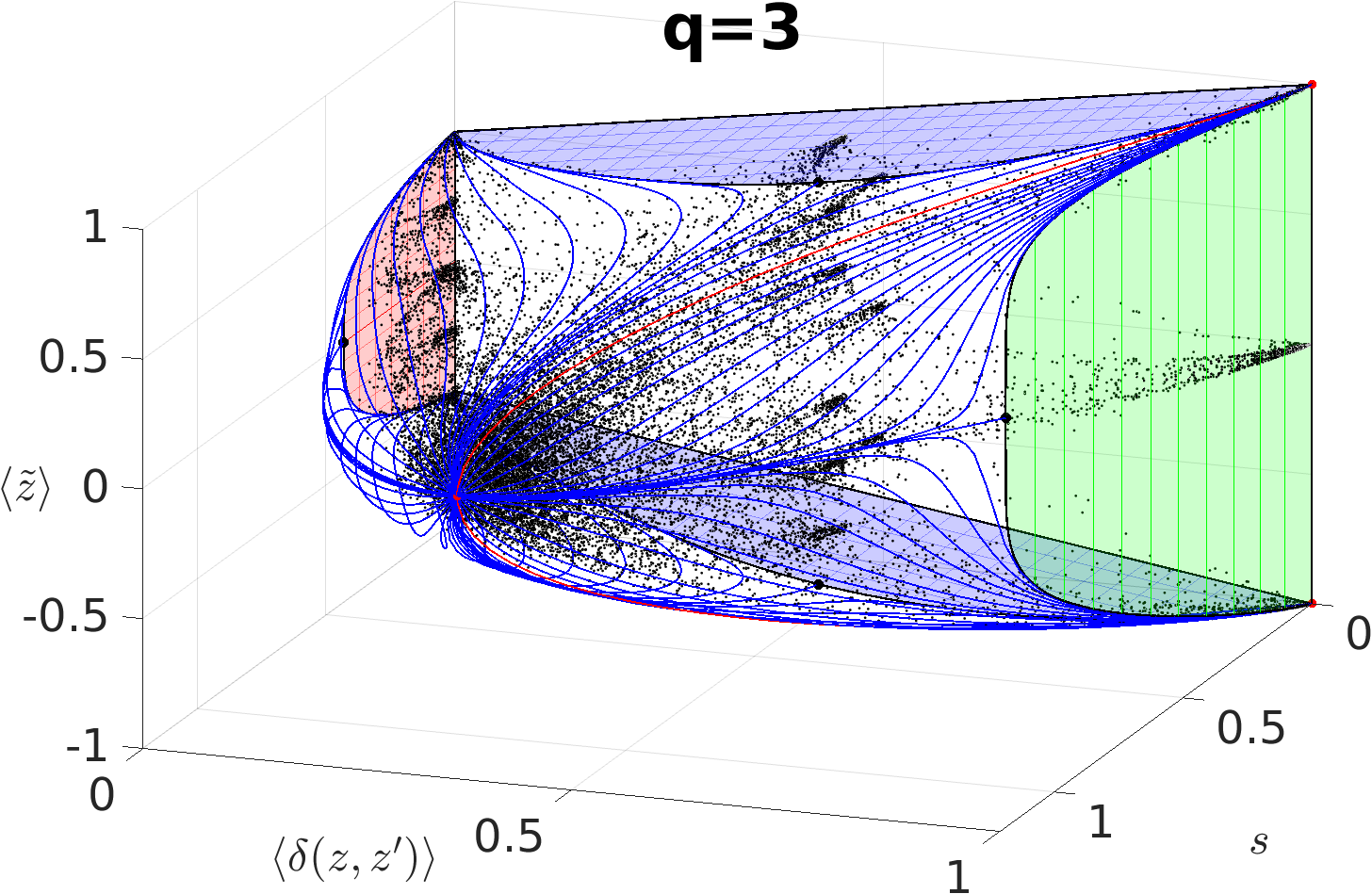}
 \includegraphics[width=0.8\linewidth,keepaspectratio=true]{\figpath/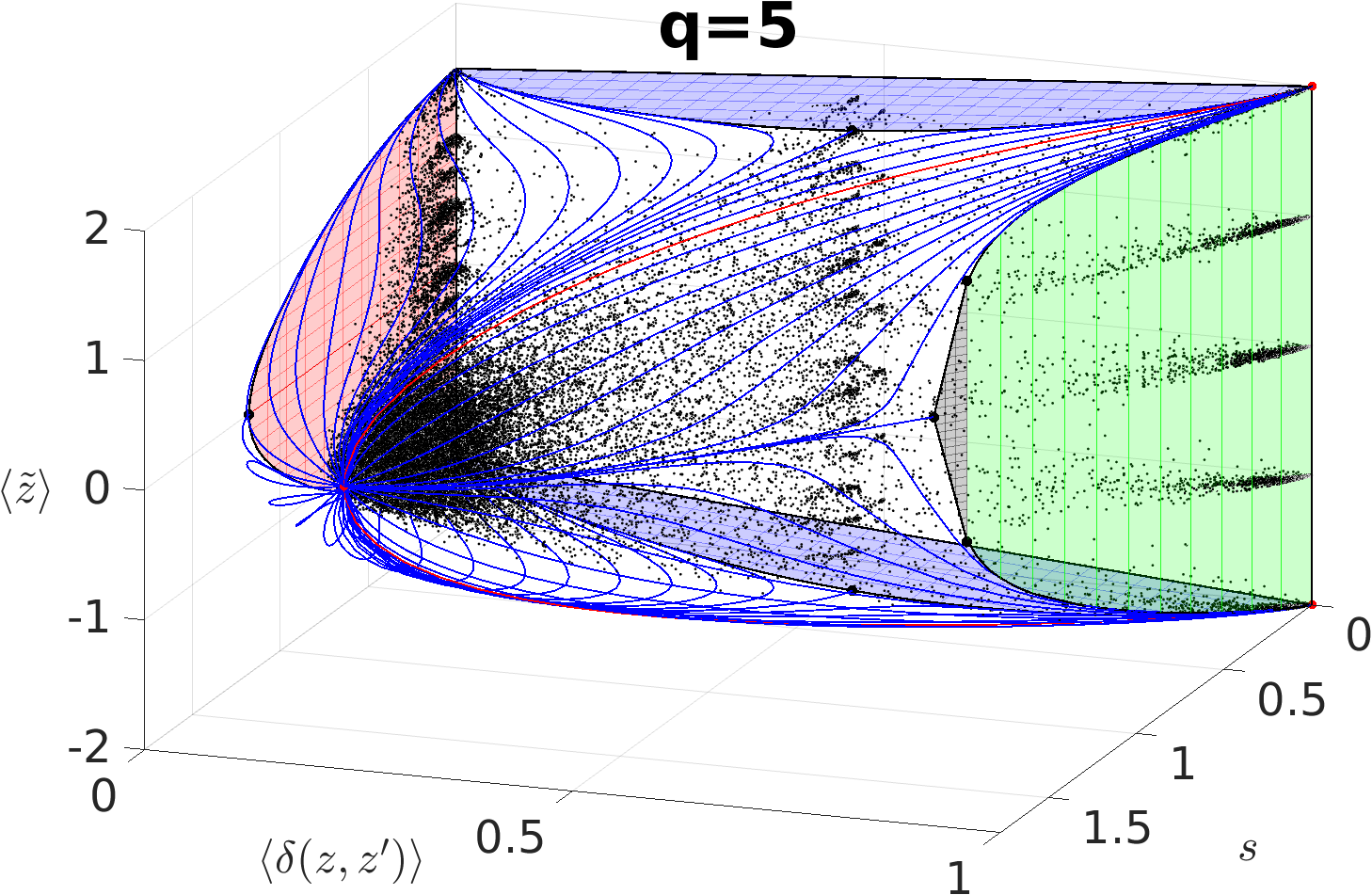}
  \caption{
 Scatter plot of observables $\edzz$, $\ezs$ and $s$ of $q$-state spins on a two-dimensional square lattice for $q=3$ and $q=5$, together with the surface of extreme points of the convex set shown in figures \ref{fig:Convex_q3} and \ref{fig:Convex_q5}. Here we explicitly plot both the upper and lower half of the convex sets for the sake of completeness. Black dots correspond to observables of single random probability distributions with interaction distance $R=1$ (see text).
 Around the green ruled surface, where there is symmetry breaking, the random scatter points clearly cluster along $q$ separate branches, whereas the other accumulation points are a finite $R$ effect.
 }
 \label{fig:scatter}
\end{figure}
In \Sec{sec:model_set} we have built on the fact that the surface of the convex sets are given by Gibbs states of \eqref{eq:PottsHam}, which can be efficiently simulated using tensor network techniques. As these convex sets however exist in probability space prior to any definition of a Hamiltonian, the occurrence of symmetry breaking is thus purely a consequence of the geometrical structure of probability space. To further elucidate this argument we show scatter plots of points from random probability distributions $P(\bz)$, not necessarily being Gibbs distributions. The points generated by expectation values with respect to these distributions must therefore all lie on or within the convex set surfaces shown in figures \ref{fig:Convex_q3} and \ref{fig:Convex_q5}.

In order to simulate random probability distributions we resort to the class of distributions representable by tensor networks consisting of 4-index tensors $T$
\begin{equation}
 P(\bz) = \prod_{n}T = T_{z_{i},z_{j},z_{k},z_{l}}T_{z_{l},z_{m},z_{n},z_{o}}T_{z_{p},z_{q},z_{j},z_{r}}T_{z_{r},z_{s},z_{m},z_{t}}\ldots,
 \label{eq:PofT}
\end{equation} 
such that the partition function, obtained by summing over all configurations $\bz$, is given by a tensor trace 
\begin{equation}
Z=\tTr\Big(\prod_{n}T\Big)=\sum_{\bz}T_{z_{i},z_{j},z_{k},z_{l}}T_{z_{l},z_{m},z_{n},z_{o}}T_{z_{p},z_{q},z_{j},z_{r}}T_{z_{r},z_{s},z_{m},z_{t}}\ldots,
\end{equation} 
which can again be efficiently evaluated using tensor network techniques. The expectation values $\edzz$ and $\ezs$ can then be calculated the usual way (c.f. \ref{sec:TNrep}).

For a general (unnormalized) probability distribution $P(\bz)$ the entropy per site is given by
\begin{equation}
 s(P) = -\sum_{\bz}P(\bz)\log[P(\bz)] = \log(z) - \frac{1}{N}\braket{\log(P)},
\end{equation} 
with $z$ the partition function per site. In the special case of a Gibbs distribution, $\braket{\log(P)}$ is nothing but the internal energy times the inverse temperature $\beta E$, which is a local observable (i.e. a sum of local terms). The entropy is then given by the familiar formula $ s = \beta(e-f)$, with $e$ the internal energy per site and $f$ the free energy per site.

For arbitrary $P(\bz)$ the quantity $\frac{1}{N}\braket{\log P}$ is in general however not a local observable.
With \eqref{eq:PofT} on the other hand we essentially restrict ourselves to probability distributions, for which the entropy is given by the sum of local observables and the entropy per site can be evaluated as
\begin{equation}
 s = \log(z) - \frac{1}{2}\braket{\log T},
 \label{eq:srand}
\end{equation} 
The factor $\frac{1}{2}$ comes from the fact that there are half as many tensors $T$ as there are sites on the lattice. We thus obtain random points within the convex set by sampling $T$ (or rather $\log T$) from some probability distribution and measuring $\edzz$, $\ezs$ and $s$ according to \eqref{eq:srand}.

The class of distributions given by \eqref{eq:PofT} contains all possible nearest-neighbor interactions as well as 4-site interactions around the face on every other plaquette. Higher order interactions and distances can be achieved in principle by blocking sites, i.e. transforming to variables $z'_{i}=\otimes_{r=1}^{R}z_{r}$. The bond dimension of $T$ is then given by $q^{R}$ and only moderate values of $R$ are computationally feasible. As a demonstration we have however resorted to $R=1$ and have drawn $\log(T)$ from a gaussian distribution with varying standard deviation $\sigma\in[0.2,1.5]$. The resulting scatter plots for $q=3$ and $q=5$ are shown in \Fig{fig:scatter}, together with the surfaces of extreme points already plotted in \fref{fig:Convex_q3} and \fref{fig:Convex_q5}.

These surfaces are asymptotically obtained by taking the convex hull of more and more random points generated that way with $R\to\infty$. \Fref{fig:scatter} shows that $R=1$ already gives a quite good qualitative approximation of the convex set. Especially around the green ruled surface where we expect spontaneous symmetry breaking it is apparent that the points cluster along $q$ distinct branches. This can be interpreted as a signature of the existence of $q$ disjoint probability spaces in the symmetry broken phase and thus statistical mixtures of configurations from different sectors do not correspond to physically realizable states. 

\section*{References}
\providecommand{\newblock}{}

\end{document}